\documentstyle[twoside,fleqn,espcrc2,epsf]{article}

\newcommand{\AmS}{{\protect\the\textfont2
  A\kern-.1667em\lower.5ex\hbox{M}\kern-.125emS}}

\title{Quenched staggered light hadron spectroscopy from
  $48^3\times64$ at $\beta = 6.5$\thanks{Poster presented by
  S.~Kim.  Computation center of RIKEN is gratefully acknowledged for
  the use of VPP-500/30.  SK is supported by KOSEF through CTP.}}

\author{Seyong Kim\address{Center for Theoretical Physics, 
        Seoul National University, Seoul, Korea}
        and 
        Shigemi Ohta\address{Institute of Particle and Nuclear Studies,
        High Energy Accelerator Research Organization (KEK), Tsukuba, Japan}}
       
\begin{document}

\begin{abstract}

We report our light hadron mass calculation based on an increased
statistics of 250 quenched gauge configurations on a \(48^3 \times
64\) lattice at \(\beta = 6.5\).  Quark propagators are calculated for
each of these configurations with staggered wall source and point sink
at quark mass values of \(m_q = 0.01, 0.005, 0.0025\) and \(0.00125\).
We also did additional calculations to improve our understanding of
systematic biases arising from autocorrelation, source size, and
propagator calculations.  Our earlier conclusions that the flavor
symmetry breaking is reduced and the ratio \(m_N/m_\rho (\sim
1.25(4))\) is small remains robust.

\end{abstract}

\maketitle

Understanding low energy properties of the strong interaction from
first principles of quantum field theory is one of the main goals for
the lattice quantum chromodynamics (QCD). Hadron spectrum is a typical
example of such low-energy phenomena and precision determination of
hadron spectrum can serve as a validity check of lattice QCD
\cite{Lat96}.

Over the past few years, we have been calculating light hadron masses
using quenched approximation to lattice QCD on a large lattice
volume (\(48^3 \times 64\)) with small lattice spacing (coupling
constant of \(\beta = 6.5\)) and small quark mass (\(m_q = 0.01,
0.005, 0.0025\) and \(0.00125\)) \cite{Ours}.  There have been
numerous efforts (see for example \cite{Lat96,GF11}) for quenched
light hadron spectrum calculation, most of which required
extrapolations with respect to lattice volume, lattice spacing, and
quark mass to approach large physical hadron size, continuum limit and
light up and down quark mass.  Controlling uncertainties arising from
such extrapolations is difficult.  In particular, extrapolations with
regard to quark mass can be troublesome due to the fact that the
chiral behavior of quenched theory is different from that of full
theory \cite{Chiral}.

We try to reduce various systematic errors associated with the
extrapolations by calculating on a large lattice volume, small lattice
spacing and small quark mass.  Previous studies suggest that for
quenched staggered spectrum, \(\beta = 6.5\) lies in asymptotic
scaling region.  Thus the question lies on whether we can simulate
large lattice volume at \(\beta=6.5\) and can reduce quark mass.

We use a combination of Metropolis and over-relaxation sweeps for
generating quenched gauge field configuration.  As we noted last year,
we increased the sweep separation between measurement from 1000 to
2000 sweeps.  This was necessary for the lightest quark mass value of
\(m_qa=0.00125\) to be free of autocorrelation, but not for the three
heavier mass values.  For hadron spectrum calculation, staggered quark
wall source with \(m_q = 0.01, 0.005, 0.0025\) and 0.00125 and point
sink is used.  For Dirac matrix inversion, standard conjugate gradient
(CG) method is used.  This year we tested the numerical robustness of
hadron propagators by comparing output from a stricter
(machine-accuracy) convergence condition. We found no significant
difference.

\begin{figure}[hbt]
\epsfxsize=70mm
\leavevmode
\epsffile[98 222 452 549]{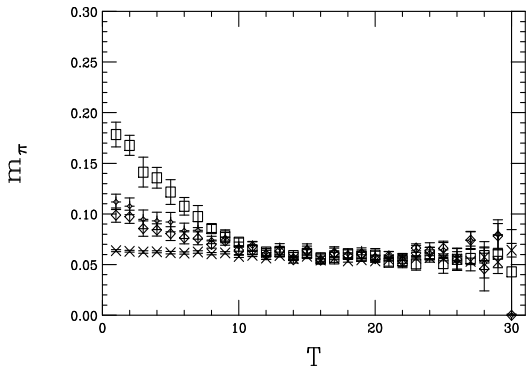}
\caption{Nambu-Goldstone pion effective mass at \protect\(\beta =
6.5\protect\) on \protect\(48^3 \times 64\protect\) lattice for 
\protect\(m_q = 0.00125\).  Three new different sizes for
corner-wall, \(12^3\) (\(\Box\)), \(24^3\) (+), \(36^3\)
(\(\Diamond\)) are used in addition to the \(48^3 (\times)\).}
\label{fig:pieff}
\end{figure}

Table \ref{tab:hadronmass} summarizes our results.  The quoted errors
are from \(\chi^2\) fit.  The corresponding jack-knife errors are in
agreement within 10 to 20 \%. We measured nucleon masses from ``corner
wall'' source and from ``even point wall'' source. However since ``corner
wall'' nucleon effective mass does not show noticeable plateau, we
present nucleon mass from ``even point wall'' source. The Goldstone 
pion mass (\(\pi\)) and non-Goldstone pion mass (\(\pi_2\)) agree
within their errors for all
the four quark mass values.  So do the two different rho mesons,
\(\rho\) and \(\rho_2\) obtained as parity partners of \(b_1\) and
\(a_1\) respectively. The flavor symmetry breaking in
these channels is still not detectable despite the refined errors.
In other words, our conclusion last year that the flavor symmetry 
breaking is not seen at \(\beta=6.5\) is robust and has been refined.
The smaller errors are obtained partly because of the increased
statistics:  the wiggling of pion effective mass also reported last
year has been reduced.  In addition we have better control in
selecting plateaus in the effective mass plot.  This became possible
through our source size study (see Figure \ref{fig:pieff}).
We clearly observe that the effective mass for all the four source
sizes eventually approach a common plateau. The unwanted contribution
from the excited states to the effective mass in earlier time
decreases as the source size is increased.  Wall size dependence for
\(m_q = 0.01\) pion effective mass is qualitatively the same.  This
behavior gives us a clear indication of how to define a plateau and
results in smaller and more reliable error estimate.  Since the
combination \(m_\pi L\) takes the values of about 2.9, 3.8, 5.3, and
7.6 respectively for quark mass \(m_qa = 0.01, 0.005, 0.0025\) and
0.00125, we do not have to worry much about finite-volume effect on
the current lattice with \(L=48\) either.

\begin{figure}[hbt]
\vspace{5pt}
\epsfxsize=60mm
\leavevmode
\epsffile[98 222 452 549]{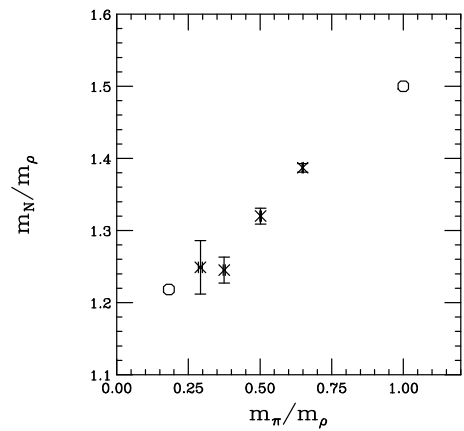}
\caption{Edinburgh plot at \protect\(\beta = 6.5\protect\) for quark mass
\protect\(m_q = 0.01\protect\), 0.005, 0.0025 and 0.00125.  Nucleon
masses from even-point wall source are used for \(m_N/m_\rho\) at each 
quark mass values.}
\label{fig:edinburgh}
\end{figure}

Figure \ref{fig:edinburgh}
shows our Edinburgh plot.  Without any extrapolation with respect to
quark mass, \(m_N/m_\rho\) approaches experimental value when the
lattice volume is large enough to admit light pion.

\begin{figure}[hbt]
\epsfxsize=60mm
\leavevmode
\epsffile[50 50 410 302]{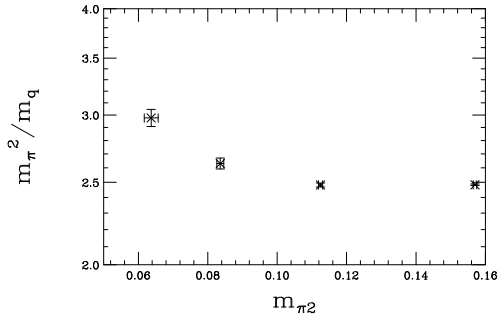}
\caption{\protect\(m_\pi^2/m_q\protect\)(vertical axis, logarithmic
scale) at \protect\(\beta = 6.5\protect\) as a function of 
\(m_{\pi_2}\)(horizontal axis).}
\label{fig:qlog}
\end{figure}

Let us look for quenched chiral logarithm in our light hadron
spectrum.  In Figure \ref{fig:qlog}, we plot \(m_\pi^2/m_q\) as a
function of \(m_{\pi_2}\).
Instead of staying flat, the data points seem to be rising, similarly
to what has been observed previously. This could be a finite volume
effect as suggested in ref.\ \cite{Bob}.
\begin{figure}[hbt]
\epsfxsize=60mm
\leavevmode
\epsffile[50 50 410 302]{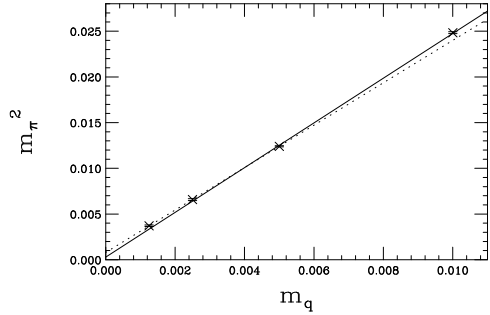}
\caption{\protect\(m_\pi^2\protect\) (vertical axis) at \protect\(\beta = 
6.5\protect\) as a function of quark mass (horizontal axis) The lines 
are results from fitting to a form, \(m_\pi^2 = c_0 + c_1 m_q\) with 
three data points only.}
\label{fig:mpisize}
\end{figure}
Therefore, in Figure \ref{fig:mpisize},
we plot \(m_\pi^2\) as a function of \(m_q\).  The two curves are
results from fitting to \(m_\pi^2 = c_0 + c_1 m_q\) following
suggestions by \cite{Bob}.  The broken line is fitted result using
\(m_q = 0.01, 0.005\) and 0.0025 with \(\chi^2/d.o.f \sim 6.7\).  The
dotted line is fitted result using \(m_q = 0.005, 0.0025\) and 0.00125
\(\chi^2/d.o.f \sim 15.4\).  Both lines show non-zero intercepts
(\(0.30(11)\times 10^{-2}\) for \(m_q = 0.01, 0.005, 0.0025\) and
\(0.80(10)\times 10^{-2}\) for \(m_q = 0.005, 0.0025,
0.00125\).  However, the slopes change noticeably from 2.45(2) for
\(m_q = 0.01, 0.005, 0.0025\) to 2.32(3) for \(m_q = 0.005, 0.0025,
0.00125\).  Since our pion mass is already small for \(m_q = 0.01\), we
think that the change in slope is less likely due to the neglected
higher order terms, \({\cal O}(m_\pi^n) (n > 2)\). Thus, our data
does not appear to agree with a finite volume cutoff plus linear term
picture of the pion mass.
\begin{table*}
\setlength{\tabcolsep}{1.0pc}
\newlength{\digitwidth} \settowidth{\digitwidth}{\rm 0}
\catcode`?=\active \def?{\kern\digitwidth}
\caption{hadron masses for \(m_q a = 0.01, 0.005, 0.0025\) 
and 0.00125}
\label{tab:hadronmass}
\begin{tabular}{cllll}
\hline
        \multicolumn{1}{c}{particle} &
        \multicolumn{1}{l}{$m_q a = 0.01$} &
        \multicolumn{1}{l}{$m_q a = 0.005$} &
        \multicolumn{1}{l}{$m_q a = 0.0025$} &
        \multicolumn{1}{l}{$m_q a = 0.00125$} \\
\hline
$\pi$    & 0.1576(3) & 0.1114(3) & 0.0811(6) & 0.0610(7) \\
$\pi_2$  & 0.1570(5) & 0.1125(6) & 0.0836(10)& 0.0637(20) \\
$\sigma$ & 0.321(3)  & 0.320(6)  & 0.340(10) & 0.291(10) \\
$\rho$   & 0.2430(7) & 0.222(1)  & 0.216(2)  & 0.209(4)  \\
$\rho_2$ & 0.2412(7) & 0.221(1)  & 0.216(2)  & 0.212(3)  \\
$a_1$    & 0.346(3)  & 0.326(4)  & 0.330(4)  & 0.311(5) \\
$b_1$    & 0.344(4)  & 0.326(6)  & 0.333(8)  &0.349(16) \\
$N$      & 0.337(1)  & 0.293(2)  & 0.269(3)  & 0.261(6)  \\
$\Delta$ & 0.397(3)  & 0.382(3)  & 0.368(5)  & 0.360(7)  \\
\hline
\end{tabular}
\end{table*}

\end{document}